\documentclass[reprint,
 amsmath,amssymb,
prb,
]{revtex4-1}

\usepackage{graphicx}
\usepackage{dcolumn}
\usepackage{longtable}
\usepackage{lipsum}
\usepackage{color}
\begin{document}


\title{ Proximity effect in a ferromagnetic semiconductor with spin-orbit interactions}
\author{Taketoki Yamashita$^{1}$}
\author{Jaechul Lee$^{1}$}
\author{Tetsuro Habe$^{1}$}
\author{Yasuhiro Asano$^{1,2,3}$}
\affiliation{$^{1}$Department of Applied Physics,
Hokkaido University, Sapporo 060-8628, Japan\\
$^{2}$Center of Topological Science and Technology,
Hokkaido University, Sapporo 060-8628, Japan\\
$^{3}$Moscow Institute of Physics and Technology, 141700 Dolgoprudny, Russia\\
}%

\date{\today}

\begin{abstract}
We study theoretically the proximity effect in a ferromagnetic semiconductor 
with Rashba spin-orbit interaction. 
The exchange potential generates opposite-spin-triplet Cooper pairs which 
are transformed into equal-spin-triplet pairs by the spin-orbit interaction. 
In the limit of strong spin-orbit interaction, symmetry of the dominant Cooper pair 
depends on the degree of disorder in a ferromagnet.
In the clean limit, spin-singlet $s$-wave Cooper pairs are the most dominant 
because the spin-momentum locking stabilizes a Cooper pair consisting of a time-reversal partner. 
In the dirty limit, on the other hand, equal-spin-triplet $s$-wave pairs are dominant  
because random impurity potentials release the locking.
We also discuss the effects of the spin-orbit interaction on the Josephson current.

\end{abstract}


\maketitle

\section{Introduction}

The proximity effect into a ferromagnetic metal has been a central issue 
in physics of superconductivity~\cite{bulaevskii:jetplett1977,buzdin:jetplett1982,bergeret:prl2001}.
The exchange potential in a ferromagnet enriches the 
symmetry variety of Copper pairs. 
The uniform exchange potential generates an opposite-spin-triplet Cooper pair
from a spin-singlet $s$-wave Cooper pair. 
The pairing function of such opposite-spin pairs oscillates and decays spatially 
in the ferromagnet, which
is a source of 0-$\pi$ transition in a superconductor/ferromagnet/superconductor
(SFS) junction~\cite{golubov:rmp2004,ryazanov:prl2001,kontos:prl2001}. 
The inhomogeneity in the magnetic moments near the junction interface 
induces equal-spin-triplet Cooper pairs which carries the long-range Josephson 
current in a SFS junction~\cite{bergeret:prl2001,keizer:nature2006,anwar:prb2010,khaire:prl2010,robinson:science2010}.
When the ferromagnet is in the diffusive transport regime, 
all the spin-triplet components belong to odd-frequency symmetry
 class.~\cite{bergeret:prl2001,asano:prl2007,braude:prl2007,eschrig:natphys2008,eschrig:rpp2015}  

An SFS junction consists of a ferromagnetic semiconductor may be a novel 
testing ground of spin-triplet Cooper pairs\cite{irie:prb2014} because of its
controllability of magnetic moment by doping. 
A long-range phase coherent effect is expected in such a high mobility two-dimensional 
electron gas on a semiconductor~\cite{takayanagi:prl1995,volkov:prl1996}.
Indeed, an experiment has observed supercurrents flowing through a Nb/(In,Fe)As/Nb 
junction~\cite{nakamura:jopcs2018,Nakamura20182}.
In addition, the spin configuration can be changed after fabricating a SFS junction 
through the Rashba spin-orbit interactions tuned by gating the ferromagnetic segment. 
It has been well established that the Rashba spin-orbit interaction generates 
the variation of spin structure in momentum space. 

So far 
the interplay between the exchange potential and the spin-orbit interaction 
in the proximity effect has been discussed in a number of theoretical studies.
\cite{demler:prb1997,buzdin:prl2008,liu:prl2014,bergeret:prl2013,bergeret:prb2014,jacobsen:prb2015-1,jacobsen:prb2015-2,costa:prb2017,mironov:prl2017}
However, symmetry of a Cooper pair contribute mainly to the Josephson current 
has never been analyzed yet in wide parameter range of the exchange potential, 
the spin-orbit potential, and the degree of disorder.
The present paper addresses this issue.

In this paper, we study theoretically the symmetries of Cooper pairs in a two-dimensional 
ferromagnetic semiconductor with the Rashba spin-orbit interaction.
The pairing function is calculated numerically by using the lattice Green's function 
technique on a SFS junction. The theoretical method can be applied to a SFS junction 
for arbitrary strength of the exchange potential, the spin-orbit interaction,
and the interactions to random impurity potential.
The pairing symmetry of the most dominant Cooper pair in a ferromagnet 
depends sensitively on the spin-orbit coupling and the degree of disorder there.
In the limit of strong spin-orbit interactions, 
a spin-singlet $s$-wave Cooper pair is dominant in a ballistic ferromagnet, whereas 
 an equal-spin-triplet $s$-wave pair is dominant in a diffusive ferromagnet.
We also discuss effects of the spin-orbit interaction on the 0-$\pi$ transition in an SFS junction.

This paper is organized as follows.
 In Sec.~II, we explain the theoretical model of an SFS junction. 
The numerical results in the clean limit and those in a dirty regime are shown in Sec.~III and IV, respectively. 
The conclusion is given in Sec.~V.
We use the units of $\hbar=c=k_B=1$ throughout this paper, where $c$ is the speed of light and 
$k_B$ is the Boltzmann constant.

\section{Model}
\begin{figure}[tbh]
  \includegraphics[width=0.45\textwidth]{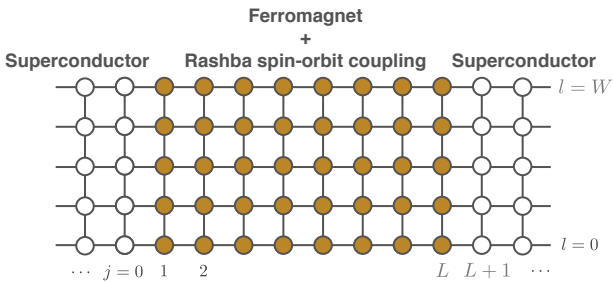}
\caption{
SFS junction on two-dimensional tight-binding model.
 }
	\label{fig:model}
\end{figure}

Let us consider an SFS junction on two-dimensional tight-binding lattice as shown in Fig.~\ref{fig:model}, 
where $L$ is the length of the ferromagnetic semiconductor, $W$ is the width of the junction 
in units of the lattice constant, $\boldsymbol{x}$ ($\boldsymbol{y}$) is the unit vector 
in the $x$ ($y$) direction, $\boldsymbol{r}=j \boldsymbol{x} + m \boldsymbol{y}$ points 
a lattice position.
The Hamiltonian of the junction is given by 
\begin{align}
\mathcal{H} =& \sum_{\boldsymbol{r}, \boldsymbol{r}^\prime}
\Psi^\dagger(\boldsymbol{r})
\left[
\begin{array}{cc}
 \hat{H}_{\mathrm{N}}(\boldsymbol{r},\boldsymbol{r}^\prime) & 
 \hat{\Delta}(\boldsymbol{r},\boldsymbol{r}^\prime) \\
 -\hat{\Delta}^\ast(\boldsymbol{r},\boldsymbol{r}^\prime) &
 -\hat{H}_{\mathrm{N}}^\ast(\boldsymbol{r},\boldsymbol{r}^\prime)
\end{array}\right]
\Psi(\boldsymbol{r}^\prime),\\
\Psi(\boldsymbol{r})=&
[\psi_{\uparrow}(\boldsymbol{r}), \psi_{\downarrow}(\boldsymbol{r}), 
\psi_{\uparrow}^\dagger(\boldsymbol{r}), \psi_{\downarrow}^\dagger(\boldsymbol{r})]^{\mathrm{T}}, 
\end{align}
where $  \psi_{\alpha}(\boldsymbol{r})$ is the annihilation operator of an electron with 
spin $\alpha$ at $\boldsymbol{r}$. 
The normal state Hamiltonian consists of four terms as,
\begin{align}
 \hat{H}_{\mathrm{N}} =& 
 \hat{H}_{\mathrm{k}} + 
 \hat{H}_{\mathrm{so}} 
 + \hat{H}_{\mathrm{h}} +\hat{V}_\mathrm{i}, \\
\hat{H}_{\mathrm{k}}(\boldsymbol{r}, \boldsymbol{r}^\prime)=& -t 
\left( \delta_{\boldsymbol{r},\boldsymbol{r}^\prime+\boldsymbol{x}} 
+ \delta_{\boldsymbol{r}+\boldsymbol{x},\boldsymbol{r}^\prime} 
\right) \hat{\sigma}_0 \nonumber\\
&-t 
\left( \delta_{\boldsymbol{r},\boldsymbol{r}^\prime+\boldsymbol{y}} 
+ \delta_{\boldsymbol{r}+\boldsymbol{y},\boldsymbol{r}^\prime} 
\right) \hat{\sigma}_0 \nonumber\\
&+(4t- \epsilon_F) \delta_{\boldsymbol{r},\boldsymbol{r}^\prime} \hat{\sigma}_0, \\
\hat{H}_{\mathrm{so}}(\boldsymbol{r}, \boldsymbol{r}^\prime)=& i (\lambda/2) 
\left[ 
\left\{ \delta_{\boldsymbol{r},\boldsymbol{r}^\prime+\boldsymbol{x}} -
 \delta_{\boldsymbol{r}+\boldsymbol{x},\boldsymbol{r}^\prime} \right\} 
 \hat{\sigma}_2 \right.\nonumber\\
-&\left.\left\{ \delta_{\boldsymbol{r},\boldsymbol{r}^\prime+\boldsymbol{y}} -
 \delta_{\boldsymbol{r}+\boldsymbol{y},\boldsymbol{r}^\prime} \right\} 
 \hat{\sigma}_1 
\right] \Theta(j) \, \Theta(L-j),\\
\hat{H}_{\mathrm{h}}(\boldsymbol{r}, \boldsymbol{r}^\prime)=& -\boldsymbol{h}\cdot \boldsymbol{\sigma} 
\, \delta_{\boldsymbol{r}, \boldsymbol{r}^\prime}\, \Theta(j)\,  \Theta(L+1-j),\\
\hat{H}_{\mathrm{i}}(\boldsymbol{r}, \boldsymbol{r}^\prime)=& v_{\boldsymbol{r}}\, \sigma_0 
\, \delta_{\boldsymbol{r}, \boldsymbol{r}^\prime}\, \Theta(j)\,  \Theta(L+1-j),\label{himp}\\
 \hat{\Delta}(\boldsymbol{r}, \boldsymbol{r}^\prime)=&
 \Delta\,  \delta_{\boldsymbol{r},\boldsymbol{r}^\prime} i\, \hat{\sigma}_2 \nonumber\\
 & \times
 \left[ \Theta(-j+1) e^{i\varphi_{L}} + \Theta(j-L) e^{i\varphi_R} \right],\\
 \Theta(j)=&\left\{ \begin{array}{cl} 1 & : j> 1\\
 0 & : j \leq 0 \end{array}\right.,
\end{align} 
where $t$ is the hopping integral among the nearest neighbor lattice sites,
$\epsilon_F$ is the Fermi energy,
$\hat{\sigma}_j$ for $j=1-3$ and $\hat{\sigma}_0$ are the Pauli's matrix and unit matrix in spin space, 
respectively. 
In the ferromagnet ($1\leq j \leq L$), $\lambda$ is the amplitude of the spin-orbit interaction, 
$\boldsymbol{h}$ represents the uniform exchange potential, and $v_{\boldsymbol{r}}$ represents 
random impurity potential.
In the two superconductors, $\Delta$ is the amplitude of the pair potential of spin-singlet $s$-wave 
symmetry, and $\varphi_{L} (\varphi_{R})$ is the superconducting phase in the left (right) 
superconductor.
The Hamiltonian is given also in continuas space in Eq.~(\ref{h-continuas}) in Appendix~A.
%
%

We solve the Gor'kov equation
\begin{align}
&\left[i\omega_n \hat{\tau}_0\hat{\sigma}_0- \sum_{\boldsymbol{r}_1}
\left(
\begin{array}{cc}
 \hat{H}_{\mathrm{N}}(\boldsymbol{r},\boldsymbol{r}_1) & 
 \hat{\Delta}(\boldsymbol{r},\boldsymbol{r}_1) \\
 -\hat{\Delta}^\ast(\boldsymbol{r},\boldsymbol{r}_1) &
 -\hat{H}_{\mathrm{N}}^\ast(\boldsymbol{r},\boldsymbol{r}_1)
\end{array}\right)
\right] \nonumber\\
&\times 
\check{G}_{\omega_n}(\boldsymbol{r}_1, \boldsymbol{r}^\prime)
= \hat{\tau}_0\hat{\sigma}_0 \delta(\boldsymbol{r}-\boldsymbol{r}^\prime),\\
&\check{G}_{\omega_n}(\boldsymbol{r}, \boldsymbol{r}^\prime)
=
\left[\begin{array}{cc}
 \hat{G}_{\omega_n}(\boldsymbol{r},\boldsymbol{r}^\prime) & 
 \hat{F}_{\omega_n}(\boldsymbol{r},\boldsymbol{r}^\prime) \\
 -\hat{F}^\ast_{\omega_n}(\boldsymbol{r},\boldsymbol{r}^\prime) &
 -\hat{G}^\ast_{\omega_n}(\boldsymbol{r},\boldsymbol{r}^\prime)
\end{array}\right],
\end{align}
by applying the lattice Green's function technique~\cite{lee:prl1981,asano:prb2001-2}, where 
$\tau_0$ is the unit matrix in particle-hole space, 
$\omega_n=(2n+1) \pi T$ is the fermionic Matsubara frequency, and $T$ is a temperature.
The Josephson current in a ferromagnet $ 1<j<L$ expressed as
\begin{align}
J(j)=& -\frac{ie}{2} \, T\sum_{\omega_n} \sum_{m=1}^{W}
\mathrm{Tr} \left[
\hat{\tau}_3 \check{T}_+\, \check{G}_{\omega_n}(\boldsymbol{r}, \boldsymbol{r}+\boldsymbol{x}) \right.\nonumber\\
& -\left.
\hat{\tau}_3 \check{T}_-\, \check{G}_{\omega_n}(\boldsymbol{r}+\boldsymbol{x}, \boldsymbol{r}) 
\right], \\
\check{T}_\pm = &\left[ \begin{array}{cc} -t \hat{\sigma}_0  \mp i  (\lambda/2) \hat{\sigma}_2 & 0 \\
0 & t \hat{\sigma}_0 \pm i  (\lambda/2) \hat{\sigma}_2
\end{array}\right],
\end{align}
 is independent of $j$.

The pairing function with $s$-wave symmetry is decomposed into 
four components
\begin{align}
\frac{1}{W}\sum_{m=1}^{W} \hat{F}_{\omega_n}(\boldsymbol{r}, \boldsymbol{r}) = \sum_{\nu=0}^3 f_{\nu}(j) 
\hat{\sigma}_\nu i\hat{\sigma}_2, \label{fp_s}
\end{align}
where $f_0$ is a spin-singlet component and $f_j$ with $j=1-3$ are spin-triplet components.
In the clean limit, we also calculate pairing function with an odd-parity symmetry 
\begin{align}
\frac{1}{2W}&\sum_{m=1}^{W} \hat{F}_{\omega_n}(\boldsymbol{r}+\boldsymbol{x}, \boldsymbol{r})
-\hat{F}_{\omega_n}(\boldsymbol{r}-\boldsymbol{x}, \boldsymbol{r}),\nonumber\\ 
&= \sum_{\nu=0}^3 f_{\nu}(j) 
\hat{\sigma}_\nu i\hat{\sigma}_2. \label{fp_o}
\end{align}

Throughout this paper, we fix several parameters as $W=20$, $\epsilon_F=2t$, $\Delta =0.005t$, and $T/T_c=0.1$.
The exchange field is always in the perpendicular direction to the two-dimensional place $\boldsymbol{h}=h \boldsymbol{z}$. 

\section{Clean limit}
\subsection{Josephson Current}
\begin{figure}[tbh]
  \includegraphics[width=0.45\textwidth]{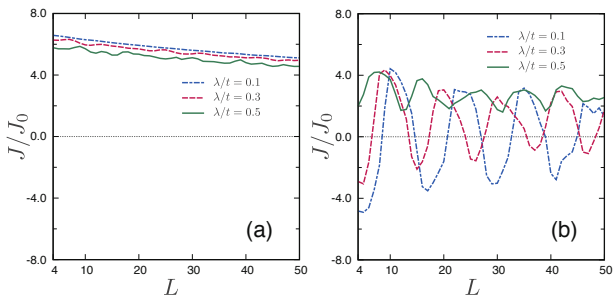}
\caption{
The Josephson current versus the length of a normal segment $L$ in the clean limit. 
(a): an SNS junction at $h=0$. (b): an SFS junction at $h=0.5t$.
 }
	\label{fig:j1c}
\end{figure}

We first discuss the numerical results of the Josephson current plotted
as a function of the length of a ferromagnet $L$ in Fig.~\ref{fig:j1c}, where we fix 
the phase difference at $\varphi=\varphi_L-\varphi_R=0.5\pi$. 
Fig.~\ref{fig:j1c}(a) and (b) show the results in the absence of exchange potential
$h=0$ and in the presence of an exchange potential $h=0.5t$, respectively. 
The Josephson current is normalized to $J_0=e\Delta$ throughout this paper.
The amplitude of the Josephson current slightly decreases with the increase of $L$ because 
the pairing functions decay as $e^{-x/\xi^C_T}$ with $\xi^C_T= v_F/ 2\pi T$ for all pairing 
symmetry. We will discuss this point later on by using analytic expression of the pairing 
function obtained by solving Eilenberger equation.
Since $h=0$ in Fig.~\ref{fig:j1c}(a), the junction corresponds to superconductor/normal-metal/superconductor 
junction. The results show that the spin-orbit interaction modifies the Josephson current very slightly.
On the other hand in Fig.~\ref{fig:j1c}(b), the Josephson current oscillates as a function of $L$ 
because of the exchange potential. The period of the oscillations is described by $\xi^C_{h}=v_F/ 2h$ 
in weak spin-orbit interactions. 
When the spin-orbit interactions increase, the amplitude of the oscillations decreases. 
At $\lambda=0.5t$, the Josephson current is always positive at $\varphi=0.5\pi$.
In the present calculation at $T/T_c=0.1$, the current-phase relationship (CPR) deviates slightly from 
sinusoidal function. Roughly speaking, 
the spin-orbit interaction stabilizes the 0 state rather than the $\pi$ state.

%
\begin{figure}[tb]
	\centering
  \includegraphics[width=0.45\textwidth]{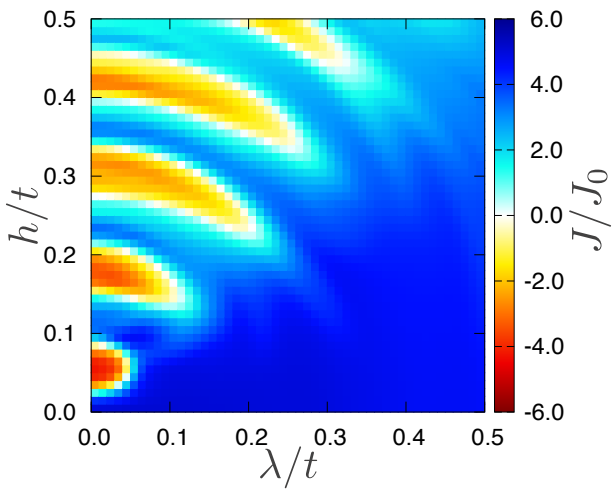}
	\caption{The Josephson current at $\varphi=0.5\pi$ and $L=50$ is plotted as a function of $h$ 
and $\lambda$ in the clean limit. 
 }
	\label{fig:phase-clean}
\end{figure}
%

In Fig.~\ref{fig:phase-clean}, we show a phase diagram of the Josephson current 
at $\varphi=0.5\pi$ and $L=50$, where horizontal (vertical) axis indicates the 
amplitude of the spin-orbit interaction (exchange potential).
The junction is in the 0 state for $J>0$ and is in the $\pi$ state for $J<0$.
At $\lambda=0$, the Josephson current changes its sign with the increase of $h$, which 
indicates the 0-$\pi$ transition by the exchange potential. 
When we introduce the spin-orbit interaction, the 0-$\pi$ transitions is suppressed 
and the Josephson current is always positive. Roughly speaking $\pi$ state 
disappears for $\lambda > h$. 
We will discuss the reasons for disappearing the $\pi$ state 
under the strong spin-orbit interactions in the next subsection.

\subsection{Pairing Functions}

To analyze the characteristic behavior of the Josephson current, 
we solve the Eilenberger equation~\cite{eilenberger:zphys1968} in a ferromagnet, 
\begin{align}
	i v_F \hat{\boldsymbol{k}}& \cdot \nabla_{\boldsymbol{r}} \check{g} +  
	[ \check{H}_0 + \check{\Delta} , \check{g} ]_{-} = 0, \\
	\check{H}_0 =& \left ( i\omega_n - \boldsymbol{h} \cdot \hat{\boldsymbol{\sigma}} \right ) \hat{\tau}_3
	- \boldsymbol{\lambda} \times \hat{\boldsymbol{\sigma}} \cdot \hat{\boldsymbol{k}}, \\
	\check{\Delta} =& i\hat{\Delta} \hat{\tau}_1.
\end{align}
To solve the Eilenberger equation, we apply the Riccati parameterization,
\begin{equation}
	\check{g} = \left( \begin{matrix}
		\hat{N} & \hat{0} \\
		\hat{0} & \underline{\hat{N}}  
	\end{matrix} \right)
	\left( \begin{matrix}
	s_{\omega_n}(1-\hat{a}\hat{\underline{a}} ) & 2 \hat{a} \\
	2 \hat{\underline{a}} & -s_{\omega_n}(1-\hat{\underline{a}}\hat{a} )	
	\end{matrix} \right),
\end{equation}
where $s_{\omega_n}=\textrm{sgn}(\omega_n)$. The two Riccati parameter
are related to each other by $\hat{\underline{a}}(\boldsymbol{r}, \hat{\boldsymbol{k}},\i\omega_n)=
\hat{\sigma}_2 \hat{a}^\ast(\boldsymbol{r},-\hat{\boldsymbol{k}},i\omega_n) \hat{\sigma}_2$.
One of the Riccati parameter obeys
\begin{align}
	i v_F &\hat{\boldsymbol{k}} \cdot \nabla \hat{a} + 2 i \omega_n \hat{a} - \boldsymbol{h} \cdot \hat{\boldsymbol{\sigma}} \hat{a} - \hat{a} \hat{\boldsymbol{\sigma}} \cdot \boldsymbol{h} - \hat{\boldsymbol{k}} \times \boldsymbol{\lambda} \cdot \hat{\boldsymbol{\sigma}} \hat{a} \nonumber\\
 &+ \hat{a} \hat{\boldsymbol{k}} \times \boldsymbol{\lambda} \cdot \hat{\boldsymbol{\sigma}} 
 - i {\Delta} + i \hat{a} {\Delta} \hat{a} = 0.
	\label{E_FR}
\end{align}

\begin{figure}[tb]
	\centering
  \includegraphics[width=0.45\textwidth]{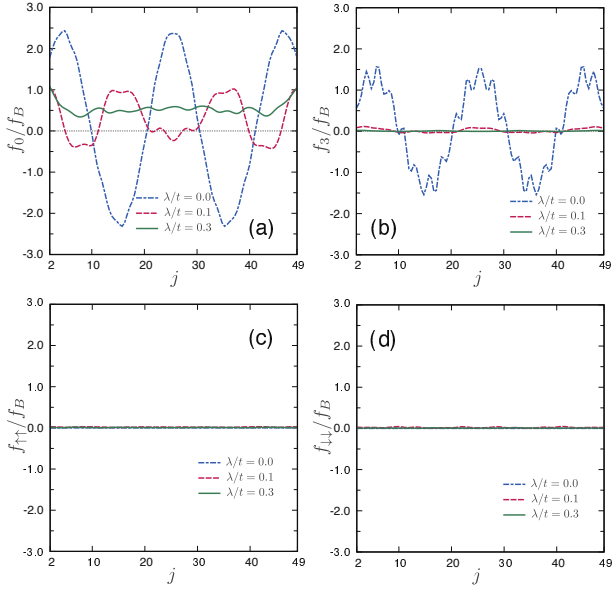}
	\caption{The spatial profile of the pairing functions at $L=50$ and $h=0.3t$ in the clean limit.
	The results for an $s$-wave symmetry in Eq.~(\ref{fp_s}) are presented.
	(a) spin-singlet $f_0$, (b) opposite-spin-triplet $f_3$, (c) equal-spin-triplet $f_{\uparrow\uparrow}$,
and (c) equal-spin-triplet $f_{\downarrow\downarrow}$ }
	\label{fig:paire-clean}
\end{figure}

\begin{figure}[tb]
	\centering
  \includegraphics[width=0.45\textwidth]{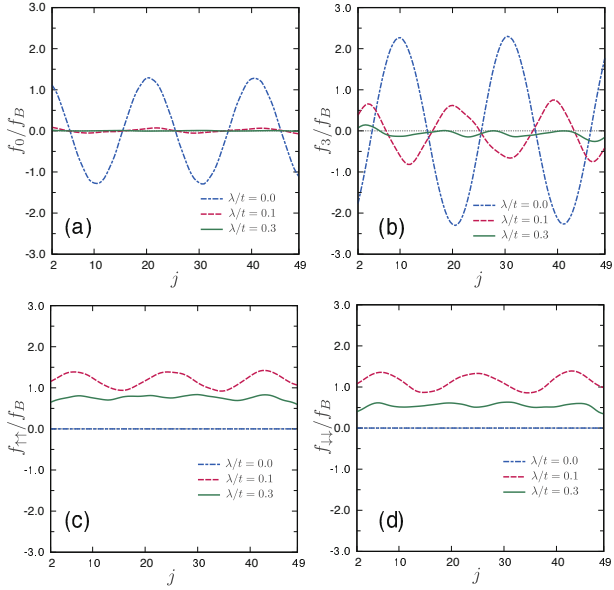}
	\caption{ The results for an odd-parity symmetry in Eq.~(\ref{fp_o}).
	The parameters are the same with those in Fig.~\ref{fig:paire-clean}.
	(a) spin-singlet $f_0$, (b) opposite-spin-triplet $f_3$, (c) equal-spin-triplet $f_{\uparrow\uparrow}$,
and (c) equal-spin-triplet $f_{\downarrow\downarrow}$ }
	\label{fig:pairo-clean}
\end{figure}

Since $\Delta=0$ in a ferromagnet ($x>0$), it is possible to have an analytic solution of
\begin{align}
\hat{a}(\boldsymbol{r}, \hat{\boldsymbol{k}},i\omega_n)= \sum_{\nu=0}^3 
a_\nu(\boldsymbol{r}, \hat{\boldsymbol{k}},i\omega_n) \hat{\sigma}_\nu. 
\end{align}
The spin-singlet component satisfies 
\begin{align}
a_0(\boldsymbol{r}, \hat{\boldsymbol{k}}, i\omega_n)= a_0(\boldsymbol{r}, -\hat{\boldsymbol{k}}, -i\omega_n),
\end{align}
and the three spin-triplet components satisfy
\begin{align}
a_j(\boldsymbol{r}, \hat{\boldsymbol{k}}, i\omega_n)= -a_j(\boldsymbol{r}, -\hat{\boldsymbol{k}}, - i\omega_n),
\end{align}
for $j=1-3$.
For $\boldsymbol{h} = h \boldsymbol{z}$ and 
$\boldsymbol{\lambda} = \lambda \boldsymbol{z}$, 
we obtain the solution in two-dimension,
\begin{align}
	a_0 =& \frac{A_0}{V^2} \left[ h^2 \cos \left( \frac{2V}{ v_{F_x}} x \right) + \lambda^2 \right]  
	e^{ -\frac{2 \omega_n}{ v_{F_x}}x}, \label{a0c}\\
	a_{\uparrow\uparrow} =& a_1 -i a_2,\\
 =& \frac{A_0 h\lambda}{V^2}(k_y + ik_x) \left[ 1- \cos \left(\frac{2V}{v_{F_x}}x \right) \right] 
 e^{ -\frac{2 \omega_n}{v_{F_x}}x},\label{auc} \\
	a_{\downarrow\downarrow} =& a_1 +i a_2,\\
 = &\frac{A_0 h\lambda}{V^2}(-k_y + ik_x)
 \left[ 1- \cos \left(\frac{2V}{v_{F_x}} x\right) \right] e^{ -\frac{2 \omega_n}{ v_{F_x}}x} ,
	\label{adc}\\
	a_3 =& i \frac{ A_0 h}{V} \sin \left( \frac{2V}{v_{F_x}} x \right) 
	e^{ -\frac{2 \omega_n}{v_{F_x}}x},
	\label{a3c}
\end{align}
where $V=\sqrt{h^2+ \lambda^2}$, $v_{F_x}= k_x/m$ and $A_0 = \Delta/(|\omega_n| + \sqrt{\omega_n^2+\Delta^2})$ is the 
solution in a uniform superconductor.
At the interface of a superconductor and a ferromagnet $(x=0)$, we imposed a boundary 
condition of $a_0=A_0$ and $a_j=0$ for $j=1-3$.
The decay length of all the components is basically given by the thermal coherence in the clean limit 
$\xi^C_T=v_F/2\pi T$. 
The spin-singlet component $a_0$ has two contributions: an oscillating term due to the exchange potential
and a constant term due to the spin-orbit interaction. 
Eqs.~(\ref{a0c})-(\ref{a3c}) suggest that only a spin-singlet pair stays in an SNS junction.
Thus the spin-orbit interaction does not affect the Josephson current so much as shown in Fig.~\ref{fig:j1c}(a). 
A opposite-spin-triplet component $a_3$ also oscillates in real space.
The pairing function for equal-spin pairs $f_{\uparrow\uparrow}$ and 
$f_{\downarrow\downarrow}$ become finite in 
the presence of the spin-orbit unteraction and oscillate in real space.
They, however, do not change their sign.

In Fig.~\ref{fig:paire-clean}, we show the numerical results of pairing function 
in the ferromagnet of an SFS junction on the tight-binding model, where $L=50$, $\varphi=0$, $h=0.3t$, 
and $\omega_n=0.02 \Delta$.
We first display the spatial profile of $s$-wave components in Eq.~(\ref{fp_s}) 
for several choices of $\lambda$.
The spin-singlet $s$-wave component $f_0$ in (a) oscillates and changes its sign in real space at $\lambda=0$.
However, the spin-orbit interaction suppresses the sign change. As a result, $f_0$ is positive at any place for $\lambda=0.3t$.
The opposite-spin-triplet component $f_3$ in (b) always changes its sing but is strongly suppressed by the 
spin-orbit interactions. 
We note that $f_3$ belongs to odd-frequency spin-triplet even-parity (OTE) symmetry class.
The two equal-spin-triplet components in (c) and (d) are absent irrespective of $\lambda$. 
The analytical results of the Eilenberger equation predict these behavior well.

In Fig.~\ref{fig:pairo-clean}, we display the spatial profile of the odd-parity 
components in Eq.~(\ref{fp_o}) for several choices of $\lambda$.
At $h=0$ and $\lambda=0$, the junction becomes an SNS junction and 
odd-parity components are absent in its normal segment. 
In the presence of the exchange potential, however, 
 odd-parity opposite-spin Cooper pairs are generated in the clean junction because 
the exchange potential breaks inversion symmetry locally at the junction interface.
The spin-singlet $s$-wave component $f_0$ in (a) oscillates and changes its sign in real 
space at $\lambda=0$.
The spin-orbit interactions suppresses drastically such an odd-frequency spin-singlet odd-parity (OSO) component.
The opposite-spin-triplet component $f_3$ in (b) belongs to 
even-frequency spin-triplet odd-parity (ETO) 
class shows similar behavior to $f_0$.
Finally, the spin-orbit interactions generate two equal-spin-triplet components $f_{\uparrow\uparrow}$ 
and $f_{\downarrow\downarrow}$ as shown in 
(c) and (d), respectively. 
They oscillate slightly in real space but do not change their sign.
The amplitude of such odd-parity equal spin components first increases with the increase of $\lambda$ 
then decrease in agreement with the analytical results in Eqs.~(\ref{auc}) and (\ref{adc}). 
The spin-momentum locking due to the spin-orbit interaction explains 
such behaviors as we discuss in what follows.

\begin{figure}[tb]
	\centering
  \includegraphics[width=0.45\textwidth]{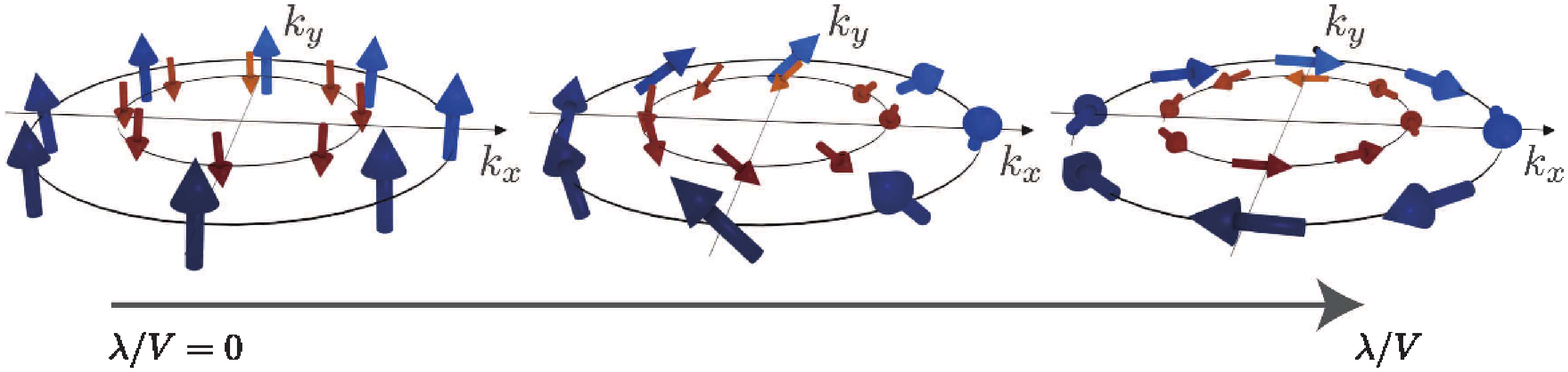}
  \includegraphics[width=0.45\textwidth]{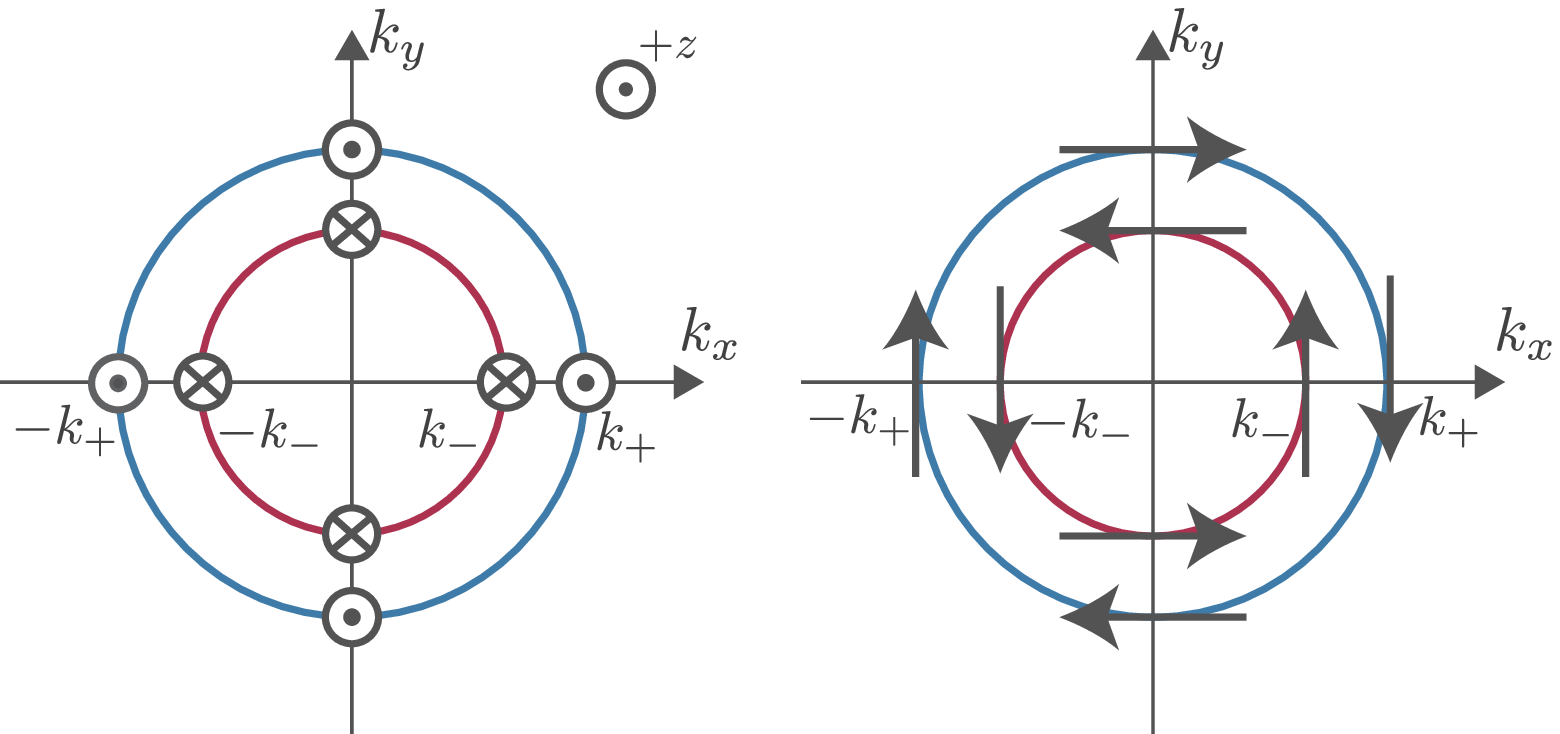}
	\caption{
	Schematic picture of spin-configuration on the Fermi surface.
	In Upper figure, the spins of an electron in the spin $\uparrow$ band are twisted 
	by the spin-orbit interactions. The strong spin-orbit interaction fells spins down to 
	a two-dimensional plane. 
	In lower figure, spin configuration on the two Fermi surface are shown for the two 
	limits: $h \gg \lambda$ on the left panel and $h \ll \lambda$ on the right panel.
	  }
	\label{fig:spin}
\end{figure}
Figure~\ref{fig:spin} shows the schematic spin structure on the Fermi surfaces.
When the exchange potential is much larger than the spin-orbit interactions, 
spin of an electron aligns in each spin bands. 
The spin-orbit interactions twists the spin structure as shown in the upper middle 
figure. The strong spin-orbit interactions causes the spin-momentum locking 
as shown in the upper right figure.
The spin configuration in the two limits are shown in the lower figure.
The wavenumber $k_\pm$ in the figure are 
$k_\pm = k_F \pm h/v_F$ in the limit of $h\gg \lambda$ on the left and 
$k_\pm = k_F \pm \lambda/v_F$ in the limit of $h\ll \lambda$ on the right.
At $h\gg \lambda$, a spin-singlet pair and 
an opposite-spin-triplet pair have 
a center-of-mass-momentum 
of $k_+ + (-k_-) =2h/v_F$ on the Fermi surface. 
As a result, these components 
oscillate and change their signs in real space~\cite{fulde_fflo,larkin_fflo}.
On the other hand, equal-spin-triplet 
components $f_{\uparrow\uparrow}$ ($f_{\downarrow\downarrow}$) 
do not have a center-of-mass-momentum because they consist 
of two electrons at $\pm k_{+}$ ($\pm k_{-}$).
Thus equal-spin-triplet components do not change signs as shown in the results 
for $\lambda=0.1t$ in Fig.~\ref{fig:pairo-clean} (c) and (d).

In the opposite limit of $h\ll \lambda$,  
a spin-singlet Cooper pair does not have a center-of-mass-momentum in this case. 
The spin-momentum locking due to strong spin-orbit interactions stabilizes such a Cooper pair 
consisting of two electrons of time-reversal partners. 
Thus a spin-singlet Cooper pair does not oscillate in real space as shown in 
the results for $\lambda=0.3t$ in Fig.~\ref{fig:paire-clean}(a).
Equal-spin-triplet pairs, on the other hand, have the center-of-mass-momentum $2\lambda/v_F$.
In Fig.~\ref{fig:pairo-clean} (c) and (d), $f_{\sigma\sigma}$ for $\lambda =0.1t$
oscillate in real space. 
Since the spacial oscillations cost the energy, $f_{\sigma\sigma}$ for $\lambda =0.3t$ 
is smaller than that for $\lambda =0.1t$. 

In the limit of $\lambda \gg h$, a spin-singlet Cooper pair is dominant in a
ferromagnet and carries the Josephson current. 
The pairing function $f_0$ in Fig.~\ref{fig:paire-clean} (a) does not changes sign.
As a result, the 0 state is stable than the $\pi$ state for $\lambda > h$ 
as shown in Fig.~\ref{fig:phase-clean}.

\section{Dirty regime}
In the dirty limit, we switch on the random impurity potential 
in Eq.~(\ref{himp}) in a ferromagnet, where
 the potential is given randomly in the range of 
$ -V_{\mathrm{imp}}/2 \leq v_{\boldsymbol{r}} V_{\mathrm{imp}}/2$. 
In the numerical simulation, we set $V_{\mathrm{imp}}=2t$, which results in 
the mean free path $\ell$ about five lattice constants. 
Since $\ell \ll L$, a ferromagnet is in the diffusive transport regime.
The coherence length $\xi_0=v_F/\pi \Delta$ is estimated as twenty lattice constants at $\Delta=0.005t$.
Thus the junction is in the dirty regime because of $\ell < \xi_0$.
The Josephson current is first calculated for a single sample with a specific 
random impurity configuration. Then the results are averaged over $N_{\mathrm{s}}$ 
samples with different impurity configurations,
\begin{align}
\langle J \rangle = \frac{1}{N_{\mathrm{s}}} \sum_{i=1}^{N_{\mathrm{s}}} J_i.
\end{align}
In this paper, we choose $N_{\mathrm{s}}$ as 100-500 in numerical simulation.

\subsection{Josephson Current}

\begin{figure}[tbh]
  \includegraphics[width=0.45\textwidth]{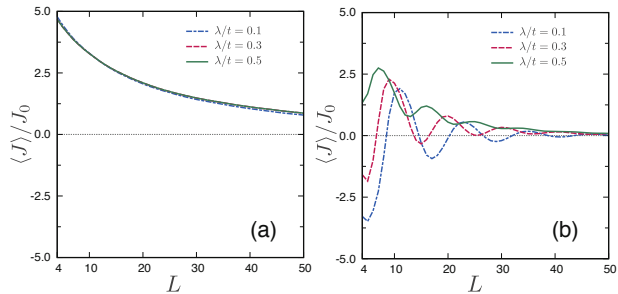}
\caption{
The Josephson current versus the length of a normal segment $L$ in the dirty limit. 
(a): an SNS junction at $h=0$. (b): an SFS junction at $h=0.5t$.
The parameters here are the same as those in Fig.~\ref{fig:j1c}
 }
 	\label{fig:j1d}
\end{figure}

In Fig.~\ref{fig:j1d}, we show the ensemble average of the Josephson current as a 
function of the length of the ferromagnet, where 
we fix $\varphi=0.5\pi$, the exchange potential is absent in (a) and the exchange potential is $h=0.5t$ in (b).
We confirmed that the current-phase relationship of ensemble aberaged Josephson current is sinusoidal.
The decay length of the Josephson current in Fig.~\ref{fig:j1d}(a) is shorter than 
that in the clean limit in Fig.~\ref{fig:j1c}(a). 
It is well known in a diffusive normal metal that, the penetration length of a Cooper pairs 
is limited by $\xi^D_{T}= \sqrt{D/2\pi T}$. The Josephson current is almost free from 
spin-orbit interactions.
The results of a SFS junction in Fig.~\ref{fig:j1d}(b) shows the oscillations 
and the sign change of the Josephson current at $\lambda=0.1t$. 
The large spin-orbit interaction suppresses the sign change of the Josephson current 
as show in the result for $\lambda=0.3t$.
%
%
\begin{figure}[tb]
	\centering
  \includegraphics[width=0.45\textwidth]{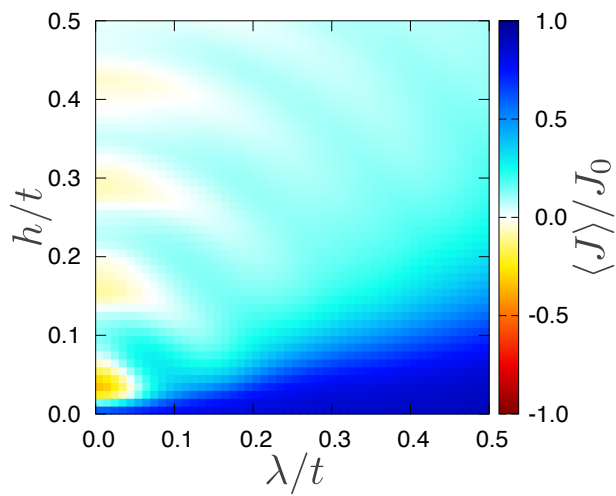}
	\caption{The Josephson current is plotted as a function of $h$ 
and $\lambda$ in the dirty regime. 
The parameters are the same as those in Fig.~\ref{fig:phase-clean}.
 }
	\label{fig:phase-dirty}
\end{figure}
%
%
In Fig.~\ref{fig:phase-dirty}, we plot the ensemble average of Josephson current 
at $\varphi=0.5\pi$ as a function of $h$ and $\lambda$.
The results should be compared with those in Fig.~\ref{fig:phase-clean}.
The amplitude of the Josephson current is suppressed by the impurity scatterings.
Although the 0-$\pi$ transition can be seen at $\lambda=0$, the spin-orbit interaction 
suppresses the 0-$\pi$ transition. 
Such tendency is common in Figs.~\ref{fig:phase-clean} and \ref{fig:phase-dirty}.

In the presence of impurities, however, the current-phase relation in a single sample 
deviates from the sinusoidal function as
\begin{align}
J_i= J_c \sin(\varphi-\varphi_i), 
\end{align}
where $\varphi_i$ is the phase shift depends on the random impurity configuration.
The numerical results for several samples are 
shown in Fig.~(\ref{fig:cpr}) in Appendix~B with broken lines, where 
at $h=0.5t$ and $\lambda=0.5t$. 
 The results show that the phase shift $\varphi_i$ depends on samples. 
The ensemble average of the results recovers the sinusoidal relationship 
as shown in a thick line in Fig.~(\ref{fig:cpr}).
The origin of the phase shift $\varphi_i$ is the 
breakdown of magnetic mirror reflection symmetry 
at the $xz$-plane by random potential. We discuss details of the symmetry breaking in Appendix~A. 
Instead of explaining magnetic mirror reflection symmetry, 
we focus on a relation between the Josephson current in theories 
and that in experiments.
In experiments, the Josephson current is measured in a specific sample of SFS junction.
Since the Josephson effect is a result of the phase coherence of a 
quasiparticle developing over a ferromagnet, the Josephson current is not a 
self-averaged quantity. 
Therefore, the Josephson current calculated at a single sample $J_i$ corresponds to that 
at a single measurement in experiments. 
When the behavior of $\langle J \rangle$ and that of $J_i$ are different 
qualitatively from each other, $\langle J \rangle$ cannot 
predict a Josephson current measured in experiments~\cite{asano:prb2001-2}.
Therefore, $\langle J \rangle$ in Fig.~\ref{fig:phase-dirty} 
tell us only a tendency of $\varphi_i$ value. 
Namely, $\varphi_i \approx 0$ would be expected in experiments for $\lambda \gg h$.
A previous paper~\cite{buzdin:prl2008} has discussed that the phase shit in 
the current-phase relationship $\varphi_0$ is tunable by 
applying the Zeeman field in the $y$ direction. 
The argument is valid only when a normal segment 
of a junction is in the ballistic transport regime 
and a junction geometry is symmetric under $y\to -y$.

\subsection{Pairing Functions}
\begin{figure}[tbh]
  \includegraphics[width=0.45\textwidth]{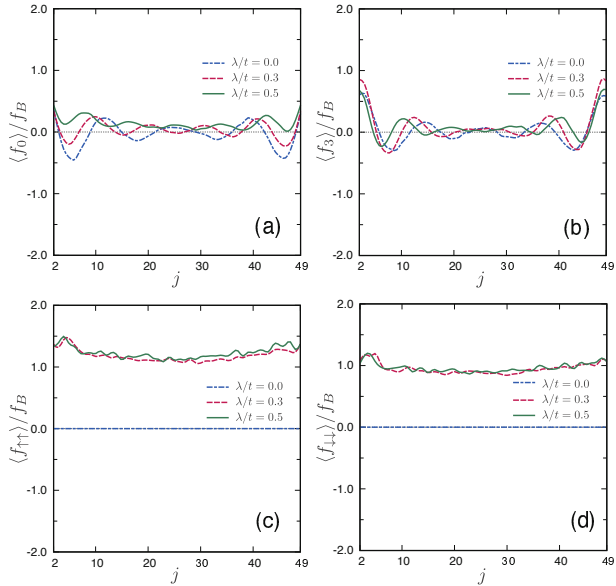}
\caption{
The spatial profile of the ensemble average of pairing function 
in the dirty regime. Only an $s$-wave component remains finite in the dirty regime.
The parameters are the same as those in Fig.~\ref{fig:paire-clean}.
 }
 	\label{fig:pair-d}
\end{figure}

Although the ensemble average of the Josephson current in theories cannot predict 
the current-phase relationship measured in a real sample, 
the ensemble average of the pairing functions tells us characteristic features of 
the proximity effect.
In Fig.~\ref{fig:pair-d}, we show the spatial profile of the pairing functions 
in dirty regime, where $\varphi=0$, $h_z=0.5t$ and $L=50$. The parameters 
here are the same as those in Fig.~\ref{fig:paire-clean}. 
The singlet component $\langle f_0 \rangle$ oscillates and changes its sign at $\lambda=0$ as
shown in Fig.~\ref{fig:pair-d}(a).
At $\lambda=0.5t$, the spin-orbit interaction surpress the amplitude of oscillations.
The similar tendency can be found also in the opposite-spin-triplet component 
$\langle f_3 \rangle$ in (b).
The equal-spin-triplet components $\langle f_{\sigma\sigma} \rangle$ are zero at $\lambda=0$.
The amplitudes of such OTE pairs become finite and spatially uniform in the dirty regime 
as shown in in Fig.~\ref{fig:pair-d}(c) and (d).
Such characteristic features in the pairing functions can be seen also in 
a single sample shown in Fig.~\ref{fig:pair-single} in Appendix~B.
Although the results in a single sample show aperiodic oscillations due to random impurity potential,
 $f_{\sigma\sigma}$ in a single sample are positive everywhere as shown in Fig.~\ref{fig:pair-single} (c) and (d).

In the clean limit, 
the spin-momentum locking suppresses the equal-spin-triplet components 
as shown in Figs.~\ref{fig:paire-clean} (c) and (d). 
In the dirty regime, however, the momentum is not a good quantum number.
Equal-spin OTE pairs have the long-range property because random impurity scatterings 
release the spin-momentum locking.\cite{bergeret:prl2013,bergeret:prb2014}
At $\lambda > g h$, the results in Fig.~\ref{fig:pair-d} show that the most dominant pair in a ferromagnet
belongs odd-frequency equal-spin-triplet $s$-wave symmetry class.

%
%
%
%

%
%
\begin{figure}[tbh]
  \includegraphics[width=0.45\textwidth]{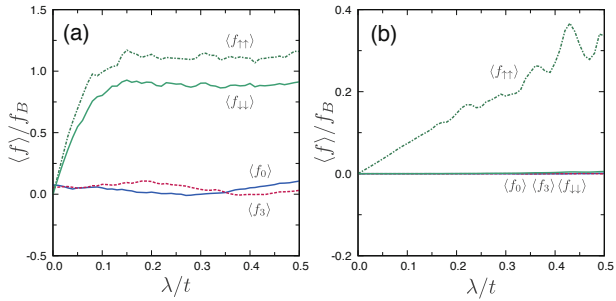}
\caption{
The pairing functions at the center of a ferromagnet $j=25$ are shown as a function of 
$\lambda$. (a): a strong ferromagnet at $h=0.5t$. (b): a half-metallic ferromagnet at $h=2.5t$.
 }
 	\label{fig:pair-dh}
\end{figure}
%

We fix $j=25$ in Fig.~\ref{fig:pair-d} at a center of a ferromagnet and calculate 
the pairing functions as a function of $\lambda$. The results are presented in 
Fig.~\ref{fig:pair-dh}, where we choose $h=0.5t$ in (a) and $h=2.5t$ in (b).
The pairing functions for opposite-spin pair $\langle f_0\rangle$ and 
$\langle f_3\rangle$ are insensitive to spin-orbit interactions. 
The amplitude of equal-spin pairs $\langle f_{\sigma\sigma} \rangle$ 
increases with 
the increase of $\lambda$ and saturate for $\lambda > 2.0t$ in (a). 
Thus odd-frequency long-range components are dominant in a strong ferromagnet.
When we increase the exchange potential to $h=2.5t$ in (b), 
a ferromagnet becomes half-metallic. 
Namely the ferromagnet is metallic for 
a spin-$\uparrow$ electron and is insulating for a spin-$\downarrow$ electron.
In such a half-metal, only $\langle f_{\uparrow\uparrow} \rangle$ component 
survives and carries the Josephson current, which is very 
similar to the situation discussed in 
the previous papers.~\cite{braude:prl2007,asano:prl2007,eschrig:natphys2008}.

\section{Conclusion}
We theoretically study the proximity effect at a ferromagnetic semiconductor 
with Rashba spin-orbit interaction by solving the Gor'kov equation on a two-dimensional 
tight-binding lattice. The Green's function is obtained numerically by using 
the lattice Green's function technique. The exchange potential in a ferromagnet converts 
a spin-singlet Cooper pair to an opposite-spin-triplet Cooper pair. The spin-orbit interactions 
generate an equal-spin-triplet Cooper pair from an opposite-spin-triplet Cooper pair. 
The relative amplitudes of the four spin pairing components 
depend on the amplitude of spin-orbit interaction and the transport regime in a ferromagnet.
In the presence of strong spin-orbit interaction,
the spin-momentum locking stabilizes a conventional 
spin-singlet $s$-wave Cooper pair in the clean limit.
In the dirty regime, on the other hand, the most dominant Cooper pair in a ferromagnet 
belongs to an odd-frequency equal-spin-triplet $s$-wave symmetry class.
The impurity scatterings release the spin-momentum locking in the dirty regime.

\begin{acknowledgments}
The authors are grateful to Ya.~Fominov, T.~Nakamura, Y.~Tanaka, and A.~A.~Golubov for useful discussions.
This work was supported by Topological Materials Science (Nos.~JP15H05852 and JP15K21717) 
from the Ministry of Education, Culture, Sports, Science and Technology (MEXT) of 
Japan, JSPS Core-to-Core Program (A. Advanced Research Networks), 
Japanese-Russian JSPS-RFBR project (Nos.~2717G8334b and 17-52-50080),
and by the Ministry of Education and Science of the Russian Federation
(Grant No.~14Y.26.31.0007).
\end{acknowledgments}

\appendix
\section{Magnetic mirror reflection symmetry}

The Hamiltonian of a SFS junction is represented inn continuous space as
\begin{align}
H(\boldsymbol{r})=&\left[ \begin{array}{cc} \xi_{\boldsymbol{r}} \hat{\sigma}_0 & \Delta_{\boldsymbol{r}} \\
-\Delta^\ast_{\boldsymbol{r}} & - \xi_{\boldsymbol{r}} \hat{\sigma}_0  \end{array}\right] 
+\left[ \begin{array}{cc}  H_{\mathrm{p}}(\boldsymbol{r}) & 0 \\
0 &  - H_{\mathrm{p}}^\ast(\boldsymbol{r}) \end{array}\right],\label{h-continuas}
\end{align}
\begin{align}
\xi_{\boldsymbol{r}} =& -\frac{\nabla^2}{2m} - \epsilon_F, \\
\Delta_{\boldsymbol{r}} =&\left[ e^{i\varphi_L}\Theta(-x) + e^{i\varphi_R}\Theta(x-L)\right] \Delta i \sigma_2,\\
H_{\mathrm{p}}=& H_{\mathrm{h}} + H_{\textrm{so}} + H_{\mathrm{i}},\\
 H_{\mathrm{h}}(\boldsymbol{r})=& -\boldsymbol{h} \cdot \hat{\boldsymbol{\sigma}}\, \Theta(x)\, \Theta(L-x), \\
 H_{\mathrm{so}}(\boldsymbol{r})= 
 &- i\lambda (\partial_y \hat{\sigma}_1 - \partial_x \hat{\sigma}_2) \Theta(x)\, \Theta(L-x),\\
 H_{\mathrm{i}}(\boldsymbol{r})=&\sum_{\boldsymbol{r}_i} v_{\boldsymbol{r}_i} \, \hat{\sigma}_0
 \delta(\boldsymbol{r}-\boldsymbol{r}_i).
\end{align}
The the eigen energy below the gap $E$ is a function on the phase difference between 
the two superconductors $\varphi-\varphi_L-\varphi_R$. 
When a relation $E(\varphi)=E(-\varphi)$ is satisfied, 
the Josephson current calculated by $J(\varphi)=e \partial_{\varphi} E$ is an odd function of $\varphi$.
In such case, the junction is either 0 or $\pi$ states, which results in $J(0)=0$.
The transformation of $\varphi \to -\varphi$ is realized by applying the complex conjugation to the Hamiltonian.
Therefore, junction is either 0 or $\pi$ state when 
\begin{align}
 H_{\mathrm{p}} =  H_{\mathrm{p}}^\ast, 
 \end{align}
is satisfied~\cite{sakurai:prb2017}. 
 In the absence of spin-orbit interactions,  
\begin{align}
 H_{\mathrm{h}} + H_{\mathrm{i}} =  \left[ H_{\mathrm{h}} + H_{\mathrm{i}} \right]^\ast,
 \end{align}
 is satisfied when the magnetic moment is spatially uniform.
It is always possible to describe the magnetic moment 
as $h_x \hat{\sigma}_1$ or $h_z \hat{\sigma}_3$ by rotating three
axes in spin space in an appropriate way.
The potentials in this paper are represented as
\begin{align}
 H_{\mathrm{p}}=&
 h_z \hat{\sigma}_3 -  i\lambda \partial_y \sigma_1 +i\lambda  \partial_x \sigma_2
 +H_{\mathrm{i}}(\boldsymbol{r}), \\
 H_{\textrm{p}}^\ast =
 & h_z \hat{\sigma}_3 +  i\lambda \partial_y \sigma_1 + i\lambda  \partial_x \sigma_2
 +H_{\mathrm{i}}(\boldsymbol{r}). 
 \end{align}
Although the second term changes it sign under the complex conjugation, 
the additional transformation $y \to -y$ cancels the sign changing,
\begin{align}
 H_{\textrm{p}}^\ast \to
 & h_z \hat{\sigma}_3 -  i\lambda \partial_y \sigma_1 + i\lambda  \partial_x \sigma_2
 +H_{\mathrm{i}}(x,-y). 
 \end{align}
Thus $E(\varphi)$ is an even function of $\varphi$ when $H_{\mathrm{i}}(x,-y)=H_{\mathrm{i}}(x,y)$ 
is satisfied.
The impurity potential is $H_{\mathrm{i}}(x,-y)\neq H_{\mathrm{i}}(x,y)$ due to its random nature.
In the presence of impurities, therefore, the energy of the junction takes its 
minimum at $\varphi=\varphi_0$ which is neither $\varphi=0$ nor $\varphi=\pi$. 
As a result, a single sample of Josephson junction with particular impurity 
configuration is 
 $\varphi_0$ junction. When we average the Josephson current 
over a number of different samples, $\langle J(\varphi) \rangle$ show the sinusoidal relation 
satisfying $\langle J(0) \rangle=0$.

A paper~\cite{buzdin:prl2008} demonstrated tuning of $\varphi_0$ by introducing the Zeeman filed 
in the $y$ direction. In this case, $H_{\mathrm{h}}^\prime = h_y \hat{\sigma}_2$ changes 
its sign under the complex conjugation. Breaking magnetic mirror reflection symmetry by $h_y$ 
explains the mechanism of the tunable feature of $\varphi_0$.

\section{Numerical results for a single sample}
In the dirty regime, calculated results for a single sample can be different from 
those of ensemble average. Here we present several results before ensemble averaging.

\begin{figure}[tbh]
  \includegraphics[width=0.4\textwidth]{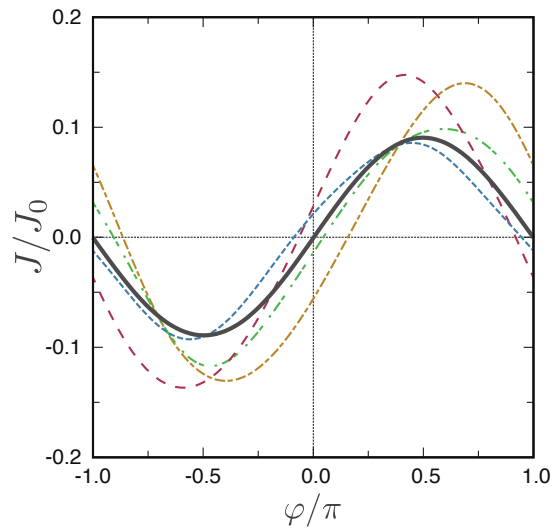}
\caption{
The current-phase relationship are shown for several samples with different 
impurity configuration at $h=0.5t$ and $\lambda=0.5t$.
The Josephson current after ensemble averaging are plotted by a thick line.
 }
 	\label{fig:cpr}
\end{figure}

In Fig.~\ref{fig:cpr}, we show the current-phase relationship 
at $h=0.5t$ and $\lambda=0.5t$. The broken lines are the results calculated for 
several samples with different random configuration and deviate the sinusoidal function. 
A thick line corresponds to the ensemble average and is sinusoidal.

%
%
%
%

\begin{figure}[tbh]
  \includegraphics[width=0.4\textwidth]{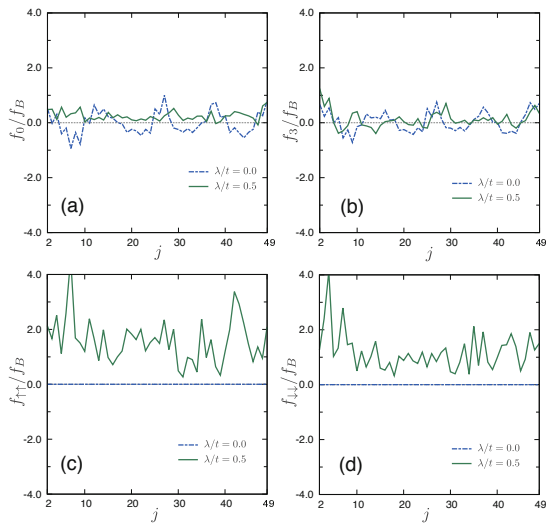}
\caption{
The spatial profile of the pairing function in a single sample.
 }
 	\label{fig:pair-single}
\end{figure}

Fig.~\ref{fig:pair-single} shows the spatial profile of the pairing function 
at $h_z=0.5t$. The results of ensemble average are shown in Fig.~\ref{fig:pair-d}
All the components oscillate aperiodically in real space due to the random 
impurity potential. 
The opposite-spin components $f_0$ and $f_3$ change their sign, whereas 
the equal-spin components $f_{\sigma,\sigma}$ do not change their sign.
Thus the results suggest the stability of equal-spin pairs in a single sample.

%
%
%
%

%


\end{document}